\documentclass[a4paper,12pt]{article}
\pdfoutput=1
\usepackage{graphicx, rotating,a4wide}
\usepackage{hyperref}
\usepackage{ifpdf,colortbl,amssymb}

\ifx\pdfoutput\undefined
\usepackage[dvips,bookmarks=false]{hyperref}	
\else
\usepackage{hyperref}	
\fi
\hypersetup{colorlinks,bookmarksopen,bookmarksnumbered,citecolor=verdes,
linkcolor=blus,pdfstartview=FitH,urlcolor=rossos}

\newcommand{\Frac}[2]{\leavevmode\kern.1em\raise.5ex
\hbox{\the\scriptfont0 #1}
\kern-.1em/\kern-.15em\lower.25ex\hbox{\the\scriptfont0 #2}}

\newcommand{\be}{\begin{equation}}
\newcommand{\ee}{\end{equation}}
\newcommand{\bea}{\begin{eqnarray}}
\newcommand{\eea}{\end{eqnarray}}
\def\eq#1{eq.~(\ref{#1})}
\def\fig#1{~\ref{fig:#1}}

\newcommand{\GeV}{\,{\rm GeV}}
\newcommand{\TeV}{\,{\rm TeV}}

\newcommand{\eV}{\,{\rm eV}}

\def\lsim{\mathrel{\rlap{\lower3pt\hbox{\hskip0pt$\sim$}}
   \raise1pt\hbox{$<$}}}         
\def\gsim{\mathrel{\rlap{\lower4pt\hbox{\hskip1pt$\sim$}}
   \raise1pt\hbox{$>$}}}         

\usepackage{multicol}
\usepackage{color}
\definecolor{rosso}{cmyk}{0,1,1,0.4}
\definecolor{rossos}{cmyk}{0,1,1,0.55}
\definecolor{rossoc}{cmyk}{0,1,1,0.2}
\definecolor{blu}{cmyk}{1,1,0,0.3}
\definecolor{blus}{cmyk}{1,1,0,0.6}
\definecolor{bluc}{cmyk}{1,1,0,0.1}
\definecolor{verde}{cmyk}{0.92,0,0.59,0.25}
\definecolor{verdec}{cmyk}{0.92,0,0.59,0.15}
\definecolor{verdes}{cmyk}{0.92,0,0.59,0.4}
\definecolor{grigio}{cmyk}{0,0,0,0.07}
\definecolor{rosa}{cmyk}{0,0.1,0.1,0.02}
\definecolor{rosino}{cmyk}{0,0.05,0.05,0.02}
\definecolor{rosas}{cmyk}{0,0.3,0.25,0.05}
\definecolor{celeste}{cmyk}{0.1,0,0,0.02}
\definecolor{giallino}{cmyk}{0,0,0.4,0.02}
\definecolor{rosso}{cmyk}{0,1,1,0.4}
\definecolor{rossos}{cmyk}{0,1,1,0.55}
\definecolor{rossoc}{cmyk}{0,1,1,0.2}
\definecolor{blu}{cmyk}{1,1,0,0.3}
\definecolor{bluc}{cmyk}{1,1,0,0.1}
\definecolor{blucc}{cmyk}{0.7,0.5,0,0}
\definecolor{viola}{cmyk}{0,1,0,0.6}
\definecolor{viola2}{cmyk}{0,1,0.2,0.6}
\definecolor{verde}{cmyk}{0.92,0,0.59,0.25}
\definecolor{verdec}{cmyk}{0.92,0,0.59,0.15}
\definecolor{verdes}{cmyk}{0.92,0,0.59,0.4}
\definecolor{verdino}{cmyk}{0.12,0,0.09,0.05}
\definecolor{giallo}{cmyk}{0,0,1,0}
\definecolor{gialloverde}{cmyk}{0.44,0,0.74,0}

\begin{document}

\color{black}

\begin{center}CERN-PH-TH/2012-098

\vspace{2cm}

{\Huge \bf\color{blus}Unificaxion}\\
\bigskip\color{black}\vspace{0.6cm}{
{\large\bf G. F. Giudice$^{a}$, R. Rattazzi$^{b}$, A. Strumia$^{c,d}$}
} \\[7mm]
{\it (a) CERN, Theory Division, CH-1211 Geneva 23, Switzerland}\\[1mm]
{\it (b) Institut de Th\'eorie des Ph\'enom\`enes Physiques, EPFL, Lausanne, Switzerland}\\[1mm]
{\it (c) Dipartimento di Fisica dell'Universit{\`a} di Pisa and INFN, Italy}\\[1mm]
{\it  (d) National Institute of Chemical Physics and Biophysics, Tallinn, Estonia}\\[1mm]
 
\end{center}
\bigskip
\bigskip
\bigskip
\vspace{1cm}

\centerline{\large\bf Abstract}

\begin{quote}\large
Dark matter, gauge coupling unification, and the strong CP problem find a common and simple solution (in the absence of naturalness) 
within  axion models. We show that such solution, even without specifying the details of the model implementation, makes testable predictions for the experimentally measurable axion parameters: the axion mass and its  coupling to photons. 
\end{quote}

\thispagestyle{empty}

\newpage


\section{Introduction}

Naturalness is considered the leading reason to believe that new physics beyond the Standard Model (SM) must exist at the weak scale. As no experimental evidence in favor of new physics at the weak scale has been observed yet, we explore the possibility of discarding the criterion of naturalness and following the lead of other arguments. Two interesting motivations to introduce new physics are Dark Matter (DM) and gauge coupling unification. Indeed these two arguments, in the absence of naturalness, have led to the intriguing hypothesis of Split Supersymmetry~\cite{split}. In this paper we will show that the same two arguments, together with the additional request of a solution to the strong CP problem, can lead to another interesting (and more minimal) hypothesis, which offers an experimentally testable prediction.

To account for DM we assume the existence of the axion, a particle which finds its motivation in the solution of the strong CP problem. Unlike the cases of the gauge hierarchy and the cosmological constant, there seems to be no anthropic explanation of the vanishingly small value of the QCD $\theta$ angle. Seeking a natural solution to the strong CP problem~\cite{axionmodels}
while giving up naturalness on the other two problems appears therefore a logical option. 
A large class of invisible axion models (the so-called KSVZ~\cite{KSVZ}) make use of new matter, charged under color and PQ symmetry, with mass at about the same scale as the axion decay constant $f_a$. It is possible that the new matter modifies the running of the SM coupling constants in such a way to achieve gauge coupling unification. We will show that this hypothesis, for which we coined the term {\it unificaxion}, leads to a prediction on the ratio between the axion coupling to photons and the the axion mass. Present and future axion experiments can test the hypothesis of unificaxion, although no new dynamics beyond the SM is predicted at the weak scale.

\medskip

This paper is organized as follows. In section~2 we discuss the requirements for new heavy fermionic particles to achieve unification of SM gauge couplings. These results are used in section~3 to predict the axion-photon coupling in unificaxion, as a function of the axion mass. We also extend our discussion to the case of supersymmetry. Finally our conclusions are drawn in section~4.

\begin{figure}[t]
$$\includegraphics[width=0.45\textwidth]{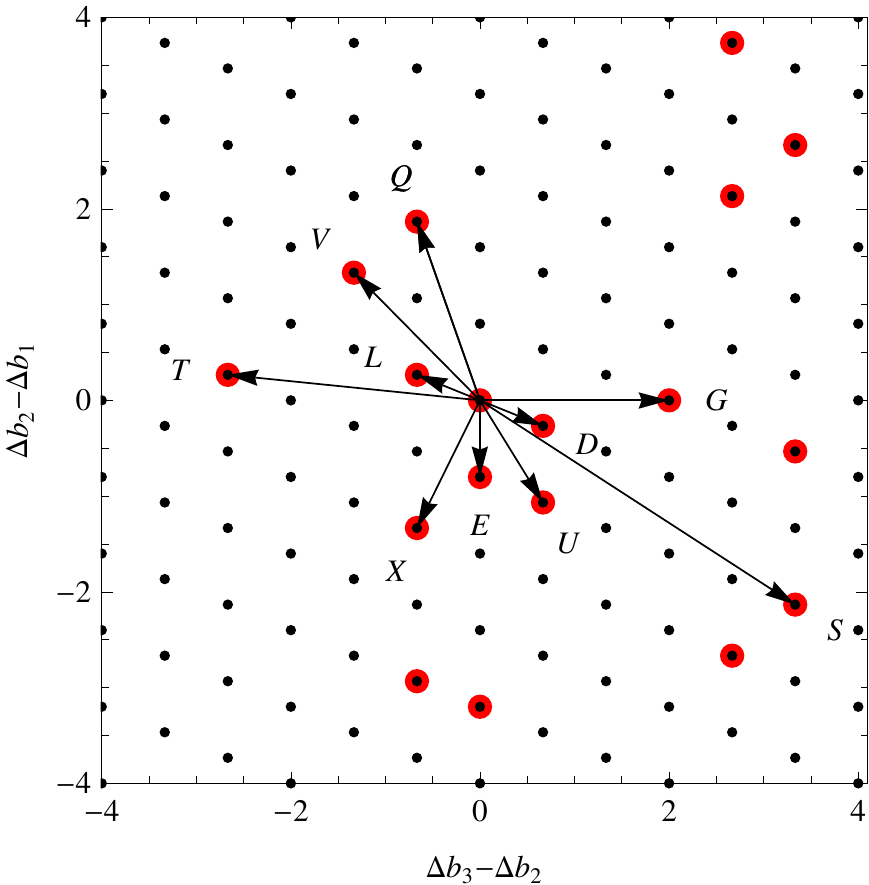}\qquad
\includegraphics[width=0.44\textwidth]{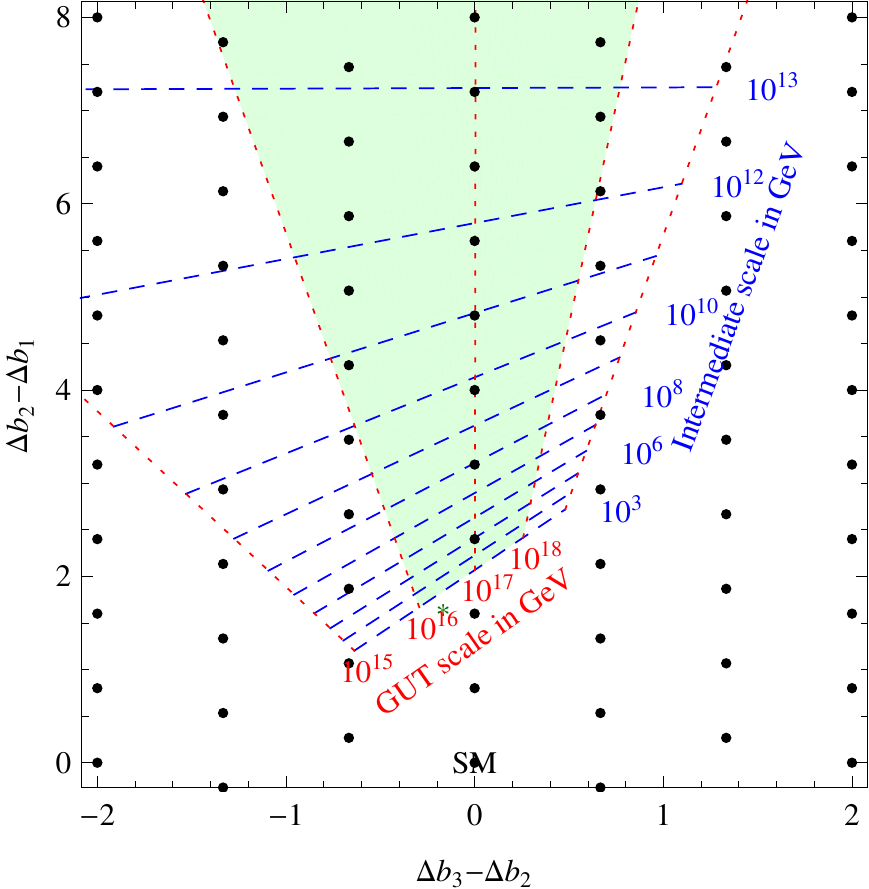}$$ 
\caption{\em {\bf Left}: The lattice formed by all possible values of the one-loop $\beta$-function coefficients generated by any combination of fermions in real representations of the SM group that can be embedded inside {\rm SU(5)} multiplets.  Thick red dots represent the contributions of single real representations, and the arrows with names represent the simplest of such cases, which are found inside the following ${\rm SU}(5)$ representations (see table~1): fundamental ($+$ conjugate)
$5\oplus\overline 5 = L \oplus D$,
antisymmetric ($+$ conjugate) $10\oplus\overline{10}=Q\oplus U\oplus D$,
symmetric ($+$ conjugate) $15\oplus\overline{15}=Q\oplus T \oplus S$, and
adjoint $24=1\oplus V\oplus G\oplus X$.
{\bf Right}:  The green area shows the range of the differences between the beta function coefficients for the gauge couplings,
$\Delta b_3 - \Delta b_2$ and $\Delta b_2 - \Delta b_1$, that provide unification
at a GUT scale between $10^{16}$ and $10^{18}\,{\rm GeV}$ (red dotted lines)
with intermediate scale indicated in blue (dashed lines). The point marked as $\star$ represents the case of low-energy supersymmetry,
and its nearest  dot represents the case of split supersymmetry with new scalars at the unification scale.
\label{fig:reticle}}
\end{figure}

\section{Unification with matter at an intermediate scale}
Let us consider the existence of new fermions $\Psi$ with common masses
$M_\Psi$. We consider fermions in real representations of the SM group (because we are interested in very massive particles) that can be embedded in SU(5) (because we have in mind a grand unified theory).
We want to explore the restrictions on the possible quantum numbers of $\Psi$   by requiring
a successful unification of the SM gauge coupling constants at some scale $M_{\rm GUT}$ in one loop approximation:
\be 
\frac{1}{\alpha_{\rm GUT}} = \frac{1}{\alpha_i(M_Z)} - \frac{b_i^{\rm SM}}{2\pi}\ln \frac{M_{\rm GUT}}{M_Z}- \frac{\Delta b_i}{2\pi}\ln \frac{M_{\rm GUT}}{M_\Psi}.
\ee
 Here $b_i^{\rm SM}
=\{41/10,-19/6,-7\} $ and $\Delta b_i$ are the contributions to the $\beta$-function coefficients
 due to SM particles and to new fermions associated with the axion sector, respectively.
  
Only the values of the differences $\Delta b_i-\Delta b_j$ are relevant for unification and for fixing its scale,
while adding a universal $\Delta b_i$ (equal for any $i$) increases the value of $\alpha_{\rm GUT}$. 
The arrows in fig.\fig{reticle}a show the values of $(\Delta b_3-\Delta b_2,\Delta b_2 - \Delta b_1)$ corresponding to real SM multiplets contained in SU(5) representations with low dimensions: $5\oplus \overline{5}$, $10\oplus \overline{10}$, $15\oplus \overline{15}$, $24$. The quantum numbers of these states are summarized in table~\ref{dinkin}. Within each complete SU(5) multiplet, arrows sum to zero.

Next, by summing these arrows with integer non-negative coefficients we generate the most generic set of points
produced by arbitrary combinations of these multiplets.
The dots represent all possible combinations: we see that they form a sparse lattice,
where each point can be produced in many different ways.
In particular, arbitrary combinations of the SM representations contained in $5\oplus\bar 5$ and $10\oplus\overline{10}$ are already
enough to span the whole lattice, and nothing more is obtained by introducing the 24 or the $15\oplus\overline{15}$.

\begin{table}\small
\begin{center}
\begin{tabular}{|cccr|cccc|c|ccc|}
\hline
${\rm SU}(5)$ & 
\multicolumn{3}{c|}{${\rm SU}(3)\otimes {\rm SU}(2) \otimes {\rm U}(1)$} & $n_3$ & ${\bar n}_3$ & $n_2$ & $z$ &name  & $\Delta b_3$ & $\Delta b_2$ & $\Delta b_1$ \\
 \hline
 \hline \rowcolor[rgb]{0.9,0.9,0.9} 
$5\oplus\bar 5$ & ~~~~~$\overline{3}$~~~~~ & 1 & $\Frac{1}{3}$~ & 0 & 1 & 0 & 0 & $D$ & 2/3 & 0  & 4/15 \\   \rowcolor[rgb]{0.9,0.9,0.9} 
$5\oplus\bar 5$&1&2&$\Frac{1}{2}$~ & 0 & 0 & 1 & 0 & $L$ & 0 & 2/3 & 2/5 \\  \rowcolor[rgb]{0.9,1,1} 
$10\oplus\overline{10}$&$\overline{3}$&1&$-\Frac{2}{3}~$& 0 & 1 & 0 & 1 & $U$ & 2/3 & 0 & 16/15\\  \rowcolor[rgb]{0.9,1,1} 
$10\oplus\overline{10}$&1&1&$-1~$& 0 & 0 & 0 & 1 & $E$ & 0 & 0 & 4/5 \\    \rowcolor[rgb]{0.9,1,1} 
$10\oplus\overline{10}$&3&2&$\Frac{1}{6}~$& 1 & 0 & 1 & 0 & $Q$ & 4/3 & 2 & 2/15 \\   \rowcolor[rgb]{1,0.9,1} 
$15\oplus\overline{15}$&3&2&$\Frac{1}{6}~$& = & = & = & = & $Q$ & =&=&= \\ \rowcolor[rgb]{1,0.9,1} 
$15\oplus\overline{15}$&1&3&$1~$& 0 & 0 & 2 & 0 & $T$ & 0 & 8/3 & 12/5\\ \rowcolor[rgb]{1,0.9,1} 
$15\oplus\overline{15}$&6&1&$-\Frac{2}{3}~$& 2 & 0 & 0 & 0 & $S$ & 10/3 & 0 & 32/15\\  \rowcolor[rgb]{1,1,0.9} 
$24$&1&3&$0~$& 0 & 0 & 2 & 1 & $V$ & 0 & 4/3 & 0\\ \rowcolor[rgb]{1,1,0.9} 
$24$&8&1&$0~$& 1 & 1 & 0 & 0 & $G$ & 2 & 0 & 0\\ \rowcolor[rgb]{1,1,0.9} 
$24$&$\overline{3}$&2&$\Frac{5}{6}~$& 0 & 1 & 1 & 0 & $X$ & 4/3 & 2 & 10/3 \\
\hline
\end{tabular}
\end{center}
\caption{\em The ${\rm SU}(3)\otimes {\rm SU}(2) \otimes {\rm U}(1)$ quantum numbers, the Dynkin labels for {\rm SU(3)} $(n_3,{\bar n}_3)$ and {\rm SU(2)} $(n_2)$, and the index $z$ for chiral irreducible representations of the SM group contained in the {\rm SU(5)} representations $5\oplus\overline 5$, $10\oplus\overline{10}$, $15\oplus\overline{15}$, $24$. The entries in the right-hand side give the contributions to $\Delta b_i$ from fermions in real representations of the SM group,
equal to the previous representations when they are real ($Y=0$),
or adding their conjugates when they are chiral ($Y\neq 0$).
\label{dinkin}}
\end{table}

\subsection{Extension to arbitrary representations}
Here we show that the lattice points shown in fig.\fig{reticle}a describe the most general case for fermionic matter and that no new points are added by including any other arbitrary irreducible representation $R$. In general, the representation $R$ can be described by the SU(3) Dynkin label ($n_3,{\bar n}_3$), the SU(2) Dynkin label ($n_2$), and hypercharge $Y$. The Dynkin labels count the differences between the number of boxes in successive rows of the corresponding Young tableau. So the indices $n_3$, ${\bar n}_3$, and $n_2$ are  non-negative integers. The contributions of $\Psi$ to the $\beta$-functions are 
\bea
\Delta b_3 &=& \frac{d}{36} \left( n_3^2+{\bar n}_3^2+n_3{\bar n}_3+3n_3+3{\bar n}_3\right) \nonumber \\
\Delta b_2 &=& \frac{d}{18} n_2(2+n_2)\label{bb1} \\
\Delta b_1 &=& \frac{2}{5}dY^2,\nonumber
\eea
where $d$ is the dimensionality of $R$, given by
\be
d=d_2d_3\qquad
d_2=1+n_2,\qquad
d_3=(1+n_3)(1+{\bar n}_3)\left(1+\frac{n_3+{\bar n}_3}{2}\right) .
\label{bb2}
\ee

The condition that $R$ is embedded in an irreducible representation of SU(5) implies a constraint on the possible values of $Y$. By projecting the weights of a generic multiplet of SU(5) into the SM subgroup we find that the hypercharge $Y$ must satisfy
\be
Y= \frac{n_2}{2} +\frac{{\bar n}_3-n_3}{3} -z.
\label{bb3}
\ee
Here $z$ is the component of a weight of the SU(5) irreducible representation, in the Dynkin basis, corresponding to the U(1) generator of the Cartan subalgebra. It can be proven that $z$ can take all possible  values in $ \mathbb{Z}$. 

As a result, $\Delta b_i$ can only scan a discrete set of values, which is determined by eq.s~(\ref{bb1})--(\ref{bb3}), with $n_2,n_3,{\bar n}_3\in \mathbb{N}$ and $z\in \mathbb{Z}$. The points obtained by this procedure are shown in fig.\fig{reticle}a as red thick circles. It is easy to see that all these points can be generated by taking appropriate combinations of the representations contained in $5\oplus\bar 5$ and $10\oplus\overline{10}$ of SU(5).

\begin{figure}[t]
$$
\includegraphics[width=0.43\textwidth]{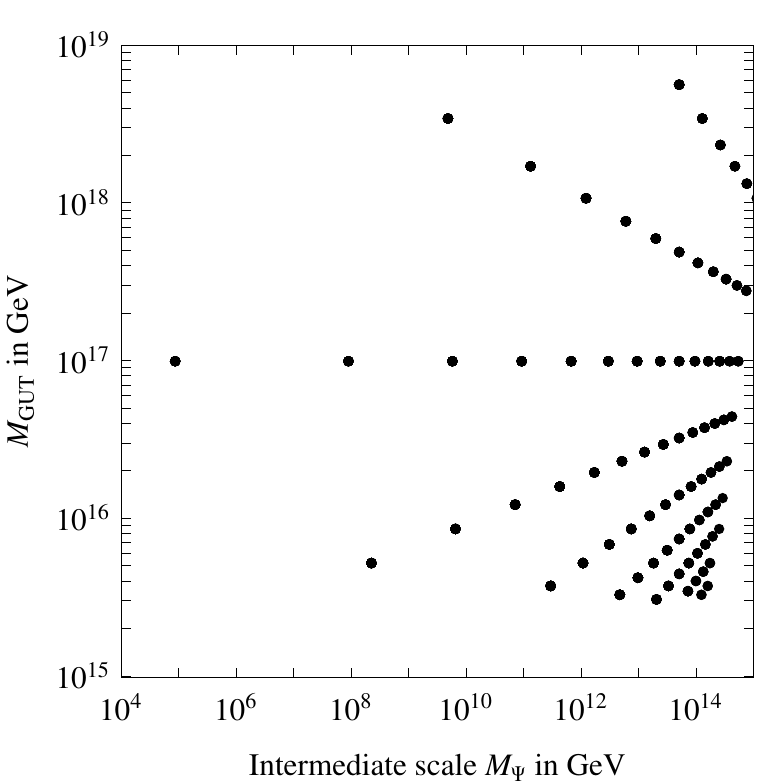}\qquad \includegraphics[width=0.5\textwidth]{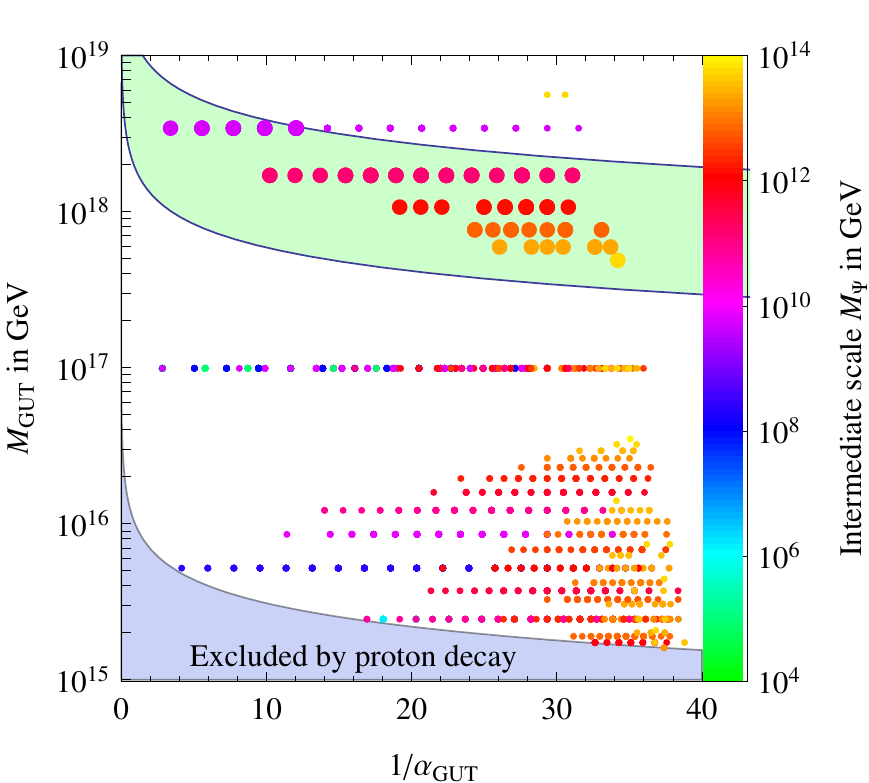}$$ 
\caption{\em {\bf Left:} The lattice points of fig.\fig{reticle} shown in terms of the intermediate scale $M_\Psi$ and the GUT scale.
{\bf Right}: The same points shown in terms of the unification coupling and mass. The color code identifies the corresponding value of the intermediate scale $M_\Psi$, as indicated.
The thick dots within the green band correspond to the range suggested by gauge/gravity unification, see eq.~(\ref{mah}).
\label{fig:GUT2}}
\end{figure}

\begin{table}
\begin{center}
\begin{tabular}{|c||c|c|c|c|}
\hline
heavy fermions & $\alpha_{\rm GUT}$ & $M_{\rm GUT}$  &  $M_\Psi$ & $E/N$\\  \hline \hline
$Q$ & 1/38 & $2\times 10^{15} \GeV $ & $1\times 10^{6} \GeV $ & 5/3\\ 
$2 Q$ & 1/38 & $2\times 10^{15} \GeV $ & $5\times 10^{10} \GeV $ & 5/3\\ 
$3 Q$ & 1/38 & $2\times 10^{15} \GeV $ & $2\times 10^{12} \GeV $ &  5/3\\ 
$ 2 Q\oplus D$ & 1/36 & $8\times 10^{15} \GeV $ & $6\times 10^{9} \GeV $ & 22/15\\ 
$2 Q \oplus U$ & 1/34 & $5\times 10^{15} \GeV $ & $2\times 10^{8} \GeV $ & 28/15\\ 
$G \oplus 2 V$ & 1/38 & $5\times 10^{15} \GeV $ & $2\times 10^{8} \GeV $ & 4/3\\  
$Q\oplus G \oplus V$ & 1/35 & $9\times 10^{16} \GeV $ & $8\times 10^{7} \GeV $ & 16/15\\ 
$Q\oplus D \oplus L $ & 1/36 & $2\times 10^{15} \GeV $ & $1\times 10^{6} \GeV $ & 2\\ 
\hline
\end{tabular}
\end{center}
\caption{\em Models of unificaxion with up to 3 fermion multiplets, intermediate mass between $10^3$ and $10^{14}\GeV$, and unification mass satisfying \eq{prot}. Their predictions for $\alpha_{\rm GUT}$, $M_{\rm GUT}$, $M_\Psi$, and $E/N$ are shown. 
\label{simplest}}
 \end{table}

\subsection{Unification}
In fig.\fig{reticle}b we show the range of $\beta$-function coefficients compatible with unification of gauge couplings
obtained by varying the intermediate mass $M_\Psi$ and the GUT scale $M_{\rm GUT}$ as indicated. The range of acceptable $M_{\rm GUT}$ is limited from below by the requirement of proton stability~\cite{SK}
\be
M_{\rm GUT}>\sqrt{\frac{\alpha_{\rm GUT}}{1/24}}~2\times 10^{15}~{\rm GeV}.
\label{prot}
\ee
An upper bound on $M_{\rm GUT}$ of the order of the Planck mass comes from the request that unification be described
at least in leading-log approximation within quantum field theory. Indeed the request $M_{\rm GUT}\lsim M_{\rm Pl}$ is also motivated
by  gauge/gravity unification.

The green dot shown in fig.\fig{reticle}b represents the one-loop prediction of low-energy supersymmetry. Note that the green dot does not exactly correspond to any of the black dots because, in addition to the fermionic degrees of freedom of gauginos and higgsinos ($G\oplus V\oplus H$), low-energy supersymmetry introduces also some new scalars in incomplete GUT multiplets (the second Higgs doublet). Notice also that the green dot sits just outside  the  region where 1-loop unification is exactly  achieved. This is a well know fact, but the mismatch is well within the size of plausible threshold corrections.

Fig.\fig{GUT2}a shows again the same lattice, this time as a function of the GUT scale and of the intermediate scale.
Of course, the discretization of these values holds up to experimental and theoretical uncertainties.
Unknown thresholds present at $M_{\rm GUT}$ and/or $M_\Psi$ are expected to be equivalent
to changing $M_{\rm GUT}$ and $M_\Psi$ by a factor of a few.
Such unknown thresholds are presumably comparable to two-loop RGE effects, which we neglect.
Furthermore, unknown non-renormalizable operators can give corrections suppressed by $M_{\rm GUT}/M_{\rm Pl}$.

Fig.\fig{GUT2}b shows the prediction for the unified coupling $\alpha_{\rm GUT}$. A wide variety of values for $\alpha_{\rm GUT}$ are possible because, for each solution of gauge coupling unification identified by the points in the region of fig.\fig{reticle}b, one can construct a tower of solutions by adding complete SU(5) representations, which do not modify $M_{\rm GUT}$, but increase $\alpha_{\rm GUT}$. The particle content of the simplest models (containing at most 3 of the fermion representations listed in table~\ref{dinkin}) are summarized in table~\ref{simplest}. These models are selected by requesting that $10^{3}\GeV <M_\Psi < 10^{14}\GeV$ and that $M_{\rm GUT}$ satisfies \eq{prot}. It is interesting that some of these models predict a value of $M_{\rm GUT}$ close to its lower value and thus predict a rate for proton decay just beyond the present experimental sensitivity.

\medskip

Furthermore, na\"{\i}ve gauge/gravity unification in 4 dimensions suggests the extra relation 
\be \alpha_{\rm GUT}
=k (\frac{M_{\rm GUT}}{M_{\rm Pl}})^2 
\label{mah}
\ee
valid up to the model-dependent order-one factor $k$.
This relation is shown as a green band in fig.\fig{GUT2}b, where we consider the range $1 < k < 40$ 
with the upper bound motivated by the heterotic string computation of~\cite{Kapl}. It should be stressed that a value 
$k\ll 1$  can be obtained in type I string theory or in M-theory as a result of a parametrically large volume
of compactification. Therefore all values of $M_{\rm GUT}\lsim M_{\rm Pl}$ are in principle compatible with the unification of all forces, though
$M_{\rm GUT}$ on the high end does seem perhaps more plausible, in that it does not require
additional very large (or very small) parameters.


\section{Axions}

In this paper we are focusing on KSVZ axion models~\cite{KSVZ} which, in addition to the SM particles, introduce a complex scalar $A$ coupled to new Dirac fermions $\Psi$, in a representation of the gauge group $\Psi =\oplus_r \psi_r$. Such models assume a Peccei-Quinn U(1) global  symmetry 
\be 
\Psi \to e^{i\gamma_5 \alpha}\Psi,\qquad A\to e^{-2i\alpha} A,
\label{postulate}
\ee
which forbids the Dirac mass term $M\,\bar\Psi \Psi$ but allows for Yukawa couplings
\be \sum_r\lambda_r\, \left (A\bar\psi_r P_L\psi_r +A^\dagger \bar\psi_r P_R\psi_r\right ) .
\label{yukawa}
\ee
The U(1) symmetry is spontaneously broken by the vacuum expectation value $\langle A\rangle = T^2 f_a$, where ${\rm Tr}\, T^a T^b = \frac{1}{2}T^2\delta^{ab}$ for the QCD generators of the fermions $\Psi$, and
 $f_a$ is the decay constant
of the light axion $a=\sqrt{2}\ {\rm Im }\, A$. We assume that all  $\lambda_r$ couplings have  comparable  size, $\lambda_r\sim \lambda_\Psi$. Consequently all fermions acquire roughly the same mass
 $M_\Psi = \lambda_\Psi \langle A\rangle$, representing the only  threshold between the weak and unification scales.
Remarkably, as we shall discuss below, the overall size of $\lambda_\Psi$ does not affect our main prediction.

\smallskip

Non-observation of axion emission from stars and supernov\ae{} implies that the axion
decay constant must be larger than $f_a> 10^9\,{\rm GeV} $.
Furthermore, requiring that the axion dark matter density generated by the initial misalignment mechanism~\cite{axionDM},
\be \Omega_a \approx 0.15  \left(\frac{f_a}{10^{12}\,\rm GeV}\right)^{7/6}\left(\frac{a_*}{f_a}\right)^2 , \ee
does not exceed the observed dark matter density $\Omega_{\rm DM} \approx 0.23$ implies the upper bound
$ f_a  < 10^{12}\,{\rm GeV}$, under the assumption that the axion vev $a_*$
in the early universe
was of the order of $f_a$~\cite{axionDM}.
Values of $f_a\sim 10^{12}\GeV$ are therefore favored by the assumption that DM is made of axions. However, larger values of $f_a$ can be compatible with axion dark matter, if $a_*$ is sufficiently smaller than its natural value of order $f_a$. 
Provided that $\lambda_\Psi = {\cal O} ( 1)$, the heavy fermions $\Psi$ that are associated with the axion dynamics have masses in the same range as $f_a$ but, for small $\lambda_\Psi$, the intermediate scale $M_\Psi$ could be much less than $10^9\,{\rm GeV} $, without conflict with axion bounds from stellar emission. Therefore, a wide range of intermediate scales $M_\Psi$ is compatible with axion DM.

\begin{figure}[t]
$$
\includegraphics[width=0.45\textwidth]{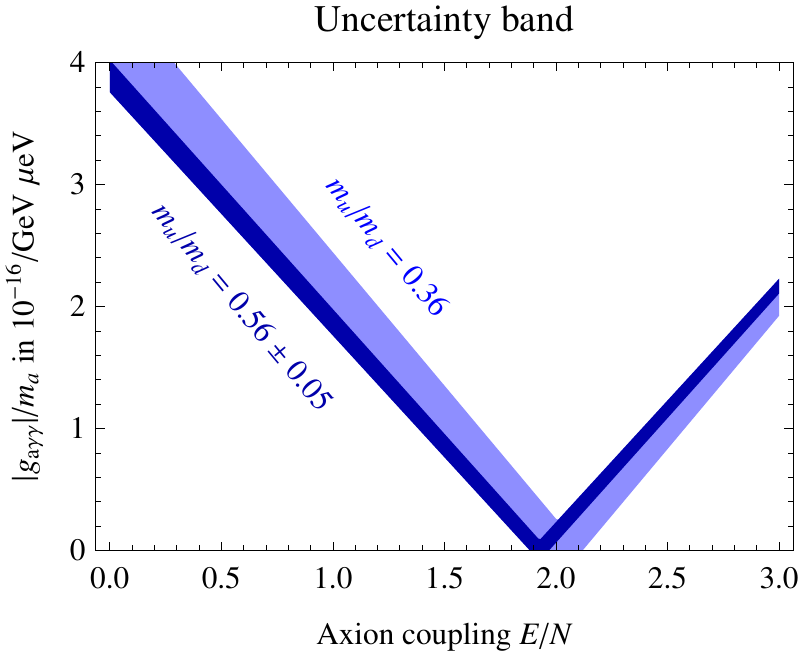}$$ 
\caption{\em The  ratio $|g_{a\gamma\gamma}|/m_a$ as a function of the model-dependent coefficient $E/N$,
taking into account the uncertainty in the quark mass ratios.
The darker band corresponds to  $m_u/m_d=0.56\pm0.05$, while the lighter band extends the uncertainty
down to $m_u/m_d=0.36$, as claimed in some analyses~\cite{PDG}. 
 \label{fig:range}}
\end{figure}

\subsection{Axion coupling to photons}
The anomalous coupling of axion to photons is defined as
\be -\frac{g_{a\gamma\gamma}}{4}  a F_{\mu\nu} \tilde F_{\mu\nu}\ee
where $\tilde F_{\mu\nu} \equiv \frac{1}{2}\epsilon_{\mu\nu\alpha\beta}F_{\alpha \beta}$.
DM axion experiments are starting to probe the theoretically favored region of 
the axion mass $m_a$ and axion coupling to photons $g_{a\gamma\gamma}$:
the  ADMX experiment~\cite{ADMX}  obtained the  limit 
\be |g_{a\gamma\gamma}| < 
\frac{7}{10^{16}\GeV} \frac{m_a}{\mu\!\eV}
\sqrt{\frac{0.3\,{\rm GeV/cm}^3}{\rho_{\rm DM}}}
\label{expt}
\ee
for $m_a$ in the range $m_a = 1.9-3.55\, \mu\!\eV$.
In \eq{expt} we have shown the dependence on the local DM density,
$\rho_{\rm DM} \approx 0.3\,{\rm GeV/cm}^3$, as unprecisely determined from rotation curves and halo dynamics.
We used the most conservative limit corresponding to completely virialized axions in the galactic halo. 
Note that, in case of discovery, the
uncertainty due to the galactic axion velocity distribution can be eliminated by studying the frequency dependence of the signal.
A positive signal would allow to precisely measure $m_a$ and determine $g_{a\gamma\gamma}$.  
The experimental determination of $m_a$ and $g_{a\gamma\gamma}$ can provide a crucial test of the idea of unificaxion, 
as we now discuss.  

\smallskip

The measurable  ratio $g_{a\gamma\gamma}/m_a$
is sensitive to the particle content of the theory:
\be\label{gagg}
\frac{g_{a\gamma\gamma}}{m_a} = \frac{\alpha_{\rm em}}{2\pi f_\pi m_\pi}\sqrt{(1+\frac{m_d}{m_u})(1+\frac{m_u}{m_d}+\frac{m_u}{m_s})}
\left[ \frac{E}{N}-\frac{2}{3}\left( \frac{4+m_u/m_d+m_u/m_s}{1+m_u/m_d+m_u/m_s}\right) \right] .
\ee
Here $f_\pi = 93$ MeV and we defined 
$E/N=\sum_r Q_{\rm PQ} q^2/\sum_r Q_{\rm PQ} T^2$ where the sums extend over all fermions $\psi_r$ with PQ charges $Q_{\rm PQ}$, electric charges $q$, and
${\rm Tr}\, T^a T^b = \frac{1}{2}T^2\delta^{ab}$ for the QCD generators. As we already explained, we are considering models in which all fermions have the same PQ charges $Q_{\rm PQ}$.
Adopting the quark masses given by chiral perturbation theory at lowest order, $m_u/m_d= 0.56$~\cite{PDG},  we get
\be 
g_{a\gamma\gamma} = \frac{2.0~(E/N - 1.92)}{10^{16}\,\rm GeV}\frac{m_a}{\mu{\rm eV}}.
\label{eq:bq}
\ee
and fig.\fig{range} shows the  band induced by the uncertainty on quark masses. 
Concerning $m_u/m_d$, we recall that second-order effects in $m_q$ in chiral perturbation theory lead to an uncertainty on the value of $m_u/m_d$ famously known as  the Kaplan-Manohar (KM) ambiguity~\cite{PDG}. Conceivably  the KM ambiguity could have  made $m_u/m_d=0$ compatible with experimental meson masses, thus disposing of the need for an axion. However that possibility is now disfavored by lattice simulations. Ref.~\cite{BM} analyzed the issue in detail also considering the impact of the KM second-order effect in eq.~(\ref{gagg}). The uncertainty band shown in fig.\fig{range} reflects the assessment of that study. It is also rather evident from eq.~(\ref{gagg}) that the leading source of uncertainty is given by the $m_d/m_u$ term in the square root factor up front, which is singular as $m_u\to 0$. In that respect, the error in  $m_s/m_u = Q\sqrt{m_d^2/m_u^2-1}$ has  a  minor impact and gives no additional uncertainty because $Q=22.7\pm0.8$ is precisely known and large.

\subsection{Unificaxion prediction for $g_{a\gamma\gamma}/m_a$}

Coming to unificaxion, we find that 
the model dependent axion coupling coefficient in eq.~(\ref{gagg}) is related to the $\beta$-function coefficients
as
\be \frac{E}{N} = \frac{\Delta b_2+5 \Delta b_1/3}{\Delta b_3}.\label{eq:qb}\ee
This result is pivotal for our analysis and so it is important to clarify the hypotheses upon which it rests. One simplifying characteristic of the class of models we are considering is, as previously mentioned, the existence of a single intermediate threshold at the energy scale $M_\Psi$.  Another condition is the existence of just one additional scalar describing the dynamics of PQ symmetry breaking and the physics at the intermediate threshold.
Besides simplicity, the lack of multiple scalar fields may be justified in the multiverse by the condition of not exacerbating the naturalness problem. Note that scalars charged under the SM gauge group would affect gauge coupling unification in a way that is completely independent of axion couplings.
Finally one last simplifying assumption, upon which our conclusions mostly rely, is that all fermions have the same PQ charge, as postulated in eq.~(\ref{postulate}). Algebraically this is the statement that the PQ charge matrix $Q_{\rm PQ}$ is proportional to the identity and, in our normalization,  $Q_{\rm PQ}= {\bf 1}$. Because of that, the anomaly coefficients  $d_{QAB}$ that control the effective coupling of the axion to the dual field strengths coincides with the contribution of fermions to the $\beta$ function coefficients
\be
d_{QAB}={\rm{Tr}}\, (Q_{\rm PQ}\{T_A,T_B\})=T^2\delta_{AB} .
\label{anomaly}
\ee
From this equation, the result in \eq{eq:qb} follows immediately.
Note that with just one scalar $A$ in principle one could accommodate fermions with PQ charge $=-1$ with Yukawa given by eq.~(\ref{yukawa}), after the replacement $A\to A^\dagger$. That would spoil the proportionality between contributions to the $\beta$ function and the axion coupling, weakening our conclusions. This could not occur in a supersymmetric theory where, in the  presence of just one scalar, holomorphy implies $Q_{\rm PQ}= {\bf 1}$. In a non-supersymmetric theory the assumption $Q_{\rm PQ}= {\bf 1}$ can be justified by assuming that all
fermions $\Psi$ sit in the same multiplet within a more fundamental description, but we will not try to construct explicit examples.
It is also worth remarking that, under the assumption $Q_{\rm PQ}= {\bf 1}$, the relation between the axion coupling and the $\beta$-function coefficients is preserved, regardless of the dynamics of the PQ breaking sector and, in particular, regardless of the number and PQ charges of the SM singlets in that sector.

\begin{figure}[t]
$$
\includegraphics[width=0.45\textwidth]{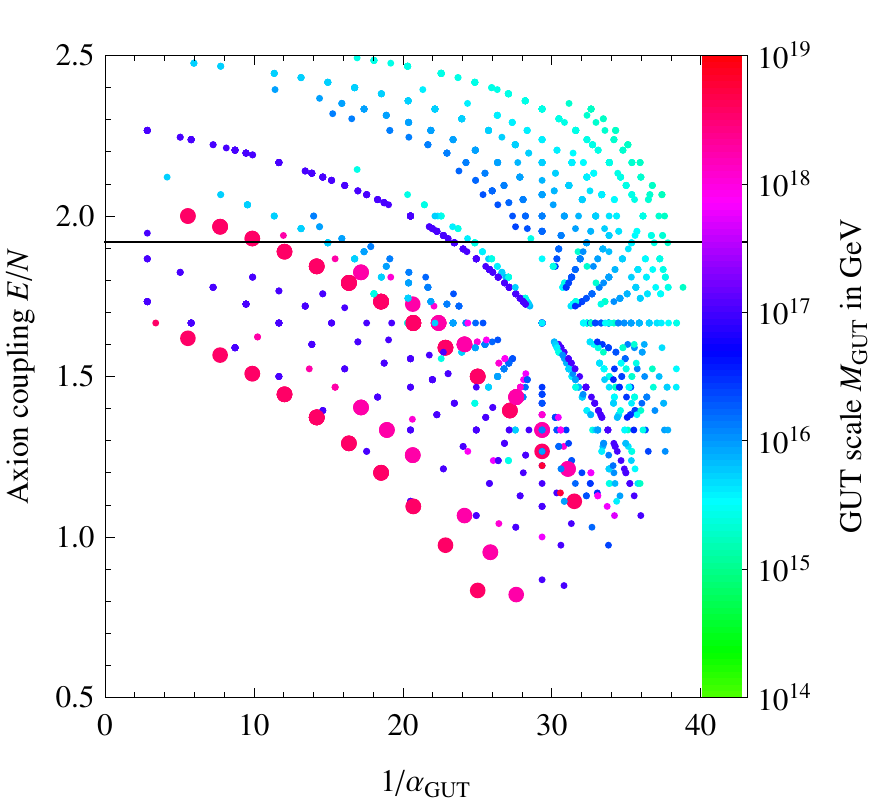}\qquad \includegraphics[width=0.45\textwidth]{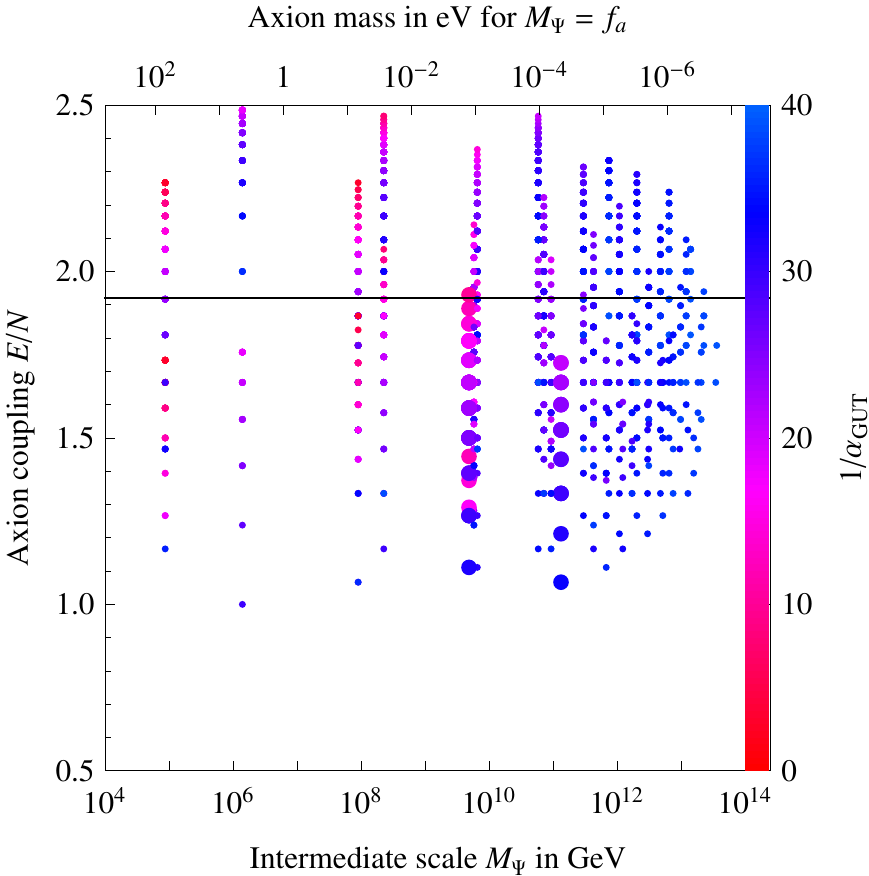}$$ 
\caption{\em {\bf Left:} 
The prediction of each unified model for $\alpha_{\rm GUT}$ 
and for $E/N$, the coefficient entering the axion coupling. The colors indicate the unified mass $M_{\rm GUT}$. The thick dots are the points identified in fig.\fig{GUT2} as suggested by gauge/gravity unification, see eq.~$(\ref{mah})$.
{\bf Right:} The same points expressed in terms of the intermediate scale
$M_\Psi$, with colors indicating the value of $\alpha_{\rm GUT}$. For guidance, we have also translated the intermediate mass $M_\Psi$ into the corresponding value of the axion mass $m_a$, under the assumption $M_\Psi =f_a$.
\label{fig:axion}}
\end{figure}

Equations (\ref{eq:bq}) and (\ref{eq:qb}) provide the link between unification and axion phenomenology, which is the key feature of unificaxion. Gauge coupling unification selects a special range for $\Delta b_i$ which, in turn, determines the measurable quantity $g_{a\gamma\gamma}/m_a$. The prediction for $g_{a\gamma\gamma}/m_a$ is obtained only by the request of unification, with no need to specify the particular particle content of the model or their interactions. 
Fig.\fig{axion} shows the correlations between $E/N$, $M_{\rm GUT}$, $\alpha_{\rm GUT}$, and $M_\Psi$. In particular, fig.\fig{axion}b illustrates how the prediction of unificaxion for $E/N$, which is directly related to $g_{a\gamma\gamma}/m_a$ through \eq{eq:bq}, depends on the intermediate mass $M_\Psi$. Under the simplifying assumption $\lambda_\Psi \approx 1$, $M_\Psi$ can be identified with $f_a$ and translated into a value of
the axion mass $m_a$ (as shown in fig.\fig{axion}b). 
However, the prediction $E/N$ is independent of $\lambda_\Psi$ and thus more robust.
  Future experimental determinations of $m_a$ and $g_{a\gamma\gamma}/m_a$ will select a region in this plane, allowing for a test of unificaxion.

It is interesting that unificaxion allows for solutions with an intermediate mass $M_\Psi$ compatible with the favored range of $f_a$, giving $1.0<E/N<2.5$. Moreover, assuming that the unified coupling is large, as maybe suggested
 in simpler models of gauge/gravity unification, we obtain $E/N> 1.6$.
 Assuming that the unification scale is very close to the Planck scale leads to
 $E/N<2.1$. A peculiar coincidence is that the predicted range of $E/N$ is centered around the value for which $g_{a\gamma\gamma}$ suffers a perfect cancellation. This is, of course, a worrisome result, because the axion would turn out to be literally invisible. Note also that $E/N$ in unificaxion is always smaller than the
 value predicted by DFSZ axion models with fermions 
 in complete SU(5) multiplets~\cite{DFSZ}, which is $E/N = 8/3$, see \eq{eq:qb}. Nevertheless, $E/N = 8/3$ gives a value of the axion-photon coupling which is inside the range predicted by unificaxion (as evident from \eq{eq:bq} and fig.\fig{range}, $|g_{a\gamma\gamma}|$ turns out to be the same for $E/N = 8/3$ and $E/N = 1.2$). Thus, the DFSZ axion or any axion model with fermions in complete GUT multiplets are experimentally indistinguishable from unificaxion, unless one devises an experiment sensitive not only to the axion coupling to photons, but also to its coupling to hadrons.

\medskip

 The ADMX bound in \eq{expt} is not yet strong enough to constrain unificaxion. However,
an improved sensitivity is considered feasible, and furthermore
new experimental techniques based on cold molecules could allow to probe also lighter axions with $f_a\sim M_{\rm GUT}$~\cite{cold}.

\begin{figure}[t]
$$\includegraphics[width=0.45\textwidth]{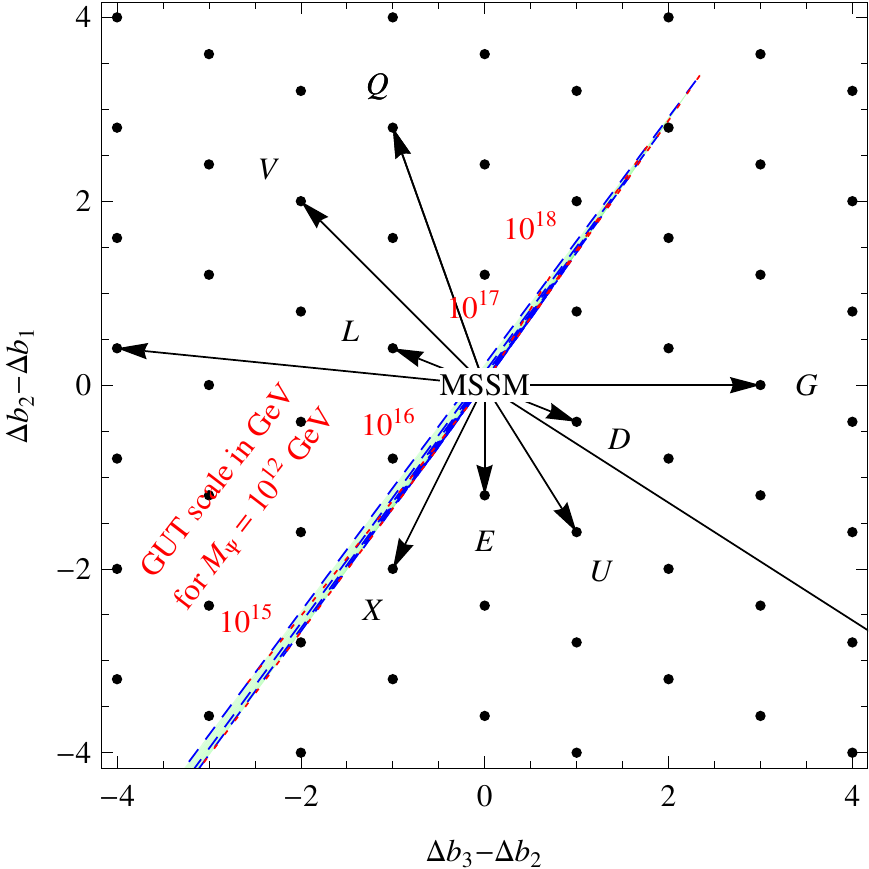}$$ 
\caption{\em The GUT lattice as in fig.\fig{reticle}, but now in the supersymmetric case.
The region that allows unification collapses to a narrow strip because
unification is already realized in low-energy supersymmetry without any extra field.
\label{fig:SUSY}}
\end{figure}

 \subsection{The supersymmetric case}
It may be of interest to extend to the case of supersymmetry our considerations about the consequences of unification for the axion coupling to photons. Let us consider low-energy supersymmetry, with new matter in chiral supermultiplets at an intermediate scale $M_\Psi$, associated with the dynamics of the  axion sector.  In this case, \eq{eq:qb} remains valid because heavy fermions and scalars
contribute to $\Delta b_i$ in the same proportion:
\be
\Delta b_i =\Delta b^F_i+\Delta b^S_i=\frac 32 \Delta b^F_i, 
\ee
where $\Delta b^{F,S}_i$ are the contributions to the $\beta$-function coefficients of the fermion and scalar components of the chiral supermultiplets.
 The only change with respect to the SM analysis amounts to replacing $b_i^{\rm SM}$ with $b_i^{\rm MSSM}=\{33/5,1,-3\}$.
 This makes however an important qualitative difference because, as well known,  in low-energy supersymmetry unification is already achieved,
 at $M_{\rm GUT}\approx 2\times10^{16}\GeV$, without the need for extra heavy multiplets.
 
As a consequence the most plausible scenario for the axion sector is to contain only complete SU(5) multiplets. In this case ($\Delta b_1=\Delta b_2=\Delta b_3$), \eq{eq:qb} gives the well-defined prediction
$E/N=8/3$. This value is larger than what expected in unificaxion but, as mentioned previously, it is indistinguishable through measurements of the effective axion-photon coupling.

However, it is also possible that new matter at the scale $M_\Psi$ modifies the gauge coupling evolution, achieving unification at a scale $M_{\rm GUT}$ different than the usual value of low-energy supersymmetry, $2\times10^{16}\GeV$. A solution of the one-loop renormalization-group equation shows that this happens when
%
\be
 \Delta b_2 -\Delta b_1 = \frac 75 \left( \Delta b_3 -\Delta b_2\right) \label{pif} \ee
 and that the new unification scale is given by
 \be \Delta b_3 -\Delta b_2 =4~\frac{\ln (M_{\rm GUT}/2\times10^{16}\GeV )}{\ln (M_{\rm GUT}/M_\Psi)},
\ee
where we  worked in the limit of exact unification for low-energy supersymmetry.
This is illustrated by the results shown in  fig.\fig{SUSY}, which is the analogous of fig.\fig{reticle}b in the case of supersymmetry. The region of $\Delta b_i$ compatible with gauge coupling unification essentially collapses to a line, approximately described by \eq{pif}.

A simple example of modified unification in low-energy supersymmetry is the addition of heavy chiral supermultiplets in the adjoint in the SM gauge group ($G\oplus V$), a case motivated by partial $N=2$ supersymmetry. In this case, we find that the intermediate scale $M_\Psi$ is related to the unification scale by
\be
M_\Psi = \left( \frac{10^{18}\GeV}{M_{\rm GUT}}\right)^3 2\times 10^{11}\GeV .
\ee
We also obtain $E/N=2/3$, a value smaller than what expected in unificaxion (and thus a larger value of $|g_{a\gamma\gamma}|$).

Another example, which gives a more accurate unification of gauge couplings, is the addition of chiral supermultiplets in the representation $G\oplus Q$. This gives
\be
M_\Psi = \left( \frac{10^{18}\GeV}{M_{\rm GUT}}\right) 4\times 10^{14}\GeV 
\ee
and, again, $E/N=2/3$.

 \section{Conclusions}
 The exploration of the `multiverse'  prompted theoretical physicists to revisit their belief in the naturalness criterion and, in some cases, to replace it with a biased statistical approach. Although interesting from a theoretical point of view, this new paradigm suffers from a chronic lack of experimental predictions. For instance, when applied to the hierarchy problem, it leads to the conclusion that it is perfectly acceptable for the Higgs boson not to be accompanied by other new particles or new dynamics at LHC energies. This conclusion appears disheartening and it can be hardly used as  evidence for the multiverse. The multiverse hypothesis is desperately in search for observational tests.

One rare exception of this lack of experimental consequences is offered by Split Supersymmetry~\cite{split}. In this context, new physics should be present at the weak scale, not because of naturalness, but because of DM and gauge coupling unification. Discovery of a long-lived gluino at the LHC would provide crucial confirmation of this hypothesis. The Higgs searches have already narrowed down the possible range of the energy scale of Split Supersymmetry~\cite{strumiag}, stating that the gluino lifetime must be $\tau_{\tilde g}<(\TeV /m_{\tilde g})^5\ 10^{-4}$ seconds. Future searches at the LHC with 14 TeV will probe the existence of a metastable gluino up to masses of about 2.5--3 TeV.

\smallskip

In this paper, we  suggested an alternative approach --- unificaxion --- for predicting new physics without invoking naturalness. The idea of unificaxion is to introduce an invisible axion model with a single complex scalar field, which spontaneously breaks a PQ abelian symmetry at a scale $f_a$ and gives masses to a set of fermions. These fermions contribute to the QCD $\theta$ term and have the appropriate quantum numbers to achieve unification of the SM gauge couplings. In this way, the strong CP problem, DM, and unification find a common solution with new physics occurring at an intermediate scale $M_\Psi$.  While the axion decay constant $f_a$ must be larger than about $10^{11}$--$10^{12}$~GeV to account for DM, the intermediate scale $M_\Psi$ of the new fermions could conceivably take even smaller values. Nonetheless, the scale $M_\Psi$ is expected to be far from the Fermi mass. Although unificaxion gives no new physics at the LHC, it makes one experimentally testable prediction in the context of axion searches. The axion-photon coupling is determined by the particle content that is responsible for gauge coupling unification. In particular, we have found that $1.0<E/N<2.5$, implying that
\be
\frac{|g_{a\gamma \gamma}|}{m_a} < 2\times 10^{-16}~ {\rm GeV}^{-1} \mu{\rm eV}^{-1}.
\label{limk}
\ee
This prediction is further narrowed down by the hypothesis that gauge unification is accompanied by an approximate gauge-gravity unification, which implies $1.6<E/N<2.1$ and thus
\be
\frac{|g_{a\gamma \gamma}|}{m_a} < 1\times 10^{-16}~ {\rm GeV}^{-1} \mu{\rm eV}^{-1}.
\ee

Present axion DM experiments are already probing an interesting range of parameters and future improvements can reach the sensitivity to test the prediction of unificaxion. 
However, it should be stressed that the discovery of an axion that satisfies \eq{limk} would only provide an indication in favor of unificaxion, but not a definitive confirmation. For example, any axion model in which the new fermions form complete GUT multiplets predicts $E/N=8/3$, leading to a value of $|g_{a\gamma \gamma}|$ inside the range of \eq{limk}. Only a test of the hadronic coupling of the axion could disentangle the two cases. Moreover,  
a worrisome feature of the unificaxion prediction is that the axion-photon coupling $g_{a\gamma \gamma}$ could be vanishingly small due to a fortuitous cancellation between short-distance and long-distance contributions. If this is the case, unificaxion would not be experimentally testable and it would remain forever buried in the obscurity of the multiverse.
 



  \small

  \paragraph{Acknowledgments}
  This work was supported by ESF grant MTT8 and by SF0690030s09 project.
We thank C. Scrucca for discussions.
The work of R.R.\ is partly supported by Swiss National Science Foundation under grant  n.\ 200020-138131.

\end{document}